\documentclass[useAMS,usenatbib]{mn2e}

\usepackage{hyperref}

\usepackage{graphicx}
\usepackage{natbib}

\bibpunct{(}{)}{;}{a}{}{,}

\providecommand{\adsurl}[1]{\href{#1}{ADS}}

\begin{document}

\title[Evolution and masses of NS and donor star in HMXB OAO~1657--415]{The evolution and masses of the neutron star and donor star in the high mass X-ray binary OAO~1657--415\thanks{Based on observations carried out at the European Southern Observatory under programme ID 081.D-0073(B)}}

\author[A.B.~Mason et al.]{A.B.~Mason$^1$\thanks{email: a.mason@open.ac.uk},  J.S.~Clark$^1$, A.J.~Norton$^1$, P.A.~Crowther$^2$,  T.M.~Tauris$^{3,4}$, N.~Langer$^{3,5}$,\newauthor I.~Negueruela$^6$, P.~Roche$^{7,8}$ \\
$^1$Department of Physics \& Astronomy, The Open University, Milton Keynes, MK7 6AA, UK. \\
$^2$Department of Physics \& Astronomy, University of Sheffield, Sheffield, S3 7RH, UK.\\
$^3$Argelander Institut f\"{u}r Astronomie, University of Bonn, Auf dem H\"{u}gel 71, D-53121, Bonn, Germany. \\
$^4$Max-Planck-Institut f\"{u}r Radioastronomie, Auf dem H\"{u}gel 69, 53121 Bonn, Germany \\
$^5$Astronomical Institute, Utrecht University, Princetonplein 5, 3584 CC, Utrecht, The Netherlands. \\
$^6$Departamento de F\'{\i}sica, Ingenier\'{\i}a de Sistemas y Teor\'{\i}a de la Se\~{n}al, Universidad de Alicante, Apdo. 99, E03080 Alicante, Spain. \\
$^7$School of Physics \& Astronomy, Cardiff University, The Parade, Cardiff, CF24 3AA, UK. \\
$^8$Division of Earth, Space \& Environment, University of Glamorgan, Pontypridd, CF37 1DL, UK.\\}

\maketitle

\begin{abstract}
We report near-infrared radial velocity measurements of the recently identified donor star in the high mass X-ray binary system OAO~1657--415 obtained in the H band using ISAAC on the VLT.

Cross-correlation methods were employed to construct a radial velocity curve with a semi-amplitude of $22.1 \pm 3.5$~km~s$^{-1}$. Combined with other measured parameters of this system this provides a dynamically determined neutron star mass of $1.42 \pm 0.26$~M$_{\odot}$ and a mass of $14.3 \pm 0.8$~M$_{\odot}$ for the Ofpe/WN9 highly evolved donor star.

OAO~1657--415 is an eclipsing High Mass X-ray binary pulsar with the largest eccentricity and orbital period of any within its class. Of the ten known eclipsing X-ray binary pulsars OAO~1657--415 becomes the ninth with a dynamically determined neutron star mass solution and only the second in an eccentric system. Furthermore, the donor star in OAO 1657--415 is much more highly evolved than the majority of the supergiant donors in other High Mass X-ray binaries (HMXBs), joining a small but growing list of HMXBs donors with extensive hydrogen depleted atmospheres.

Considering the evolutionary development of OAO~1657--415, we have estimated the binding energy of the envelope of the mass donor and find that there is insufficient energy for the removal of the donor's envelope via spiral-in, ruling out a Common Envelope evolutionary scenario. With its non-zero eccentricity and relatively large orbital period the identification of a definitive evolutionary pathway for OAO 1657--415 remains problematic, we conclude by proposing two scenarios which may account for OAO 1657--415 current orbital configuration.

\end{abstract}

\begin{keywords}
binaries:eclipsing - binaries:general - X-rays:binaries - stars:individual:OAO~1657--415 - stars:massive - stars:Wolf-Rayet
\end{keywords}

\section{Introduction}

\subsection{Neutron star masses}

Despite much theoretical work and the proposal of over 100 different neutron star (NS) equations of state (EoS) \citep{kaper06b}, the precise form of the fundamental physical properties
of matter under the extreme densities and pressures found within NS are still unknown. In this regard, observational data can help to eliminate contending EoS by rejecting those that place unreasonable constraints on the mass range of observed NS. The recent measurement of 1.97$\pm$0.04~M$_{\odot}$ for
the mass of the binary millisecond pulsar (MSP) PSR~J1614-2230  already excludes many EoS invoking the presence of exotic hadronic matter \citep{demorest10}, while higher NS masses - albeit
with greater uncertainty - have also been reported (see Section 6).
Moreover, one might expect the mass of NSs to show a dependence on the prior evolutionary history, such that higher masses might be expected in binaries where significant accretion
has occurred (e.g. binary  MSPs) or where  massive progenitors have yielded a high pre-SN core mass. With regard to the former,  \citet{zhang10} have recently
proposed that the mean mass of MSPs is greater than that of the total NS population, while \citet{timmes96} propose a bimodal distribution of NS masses (peaking at 1.27 \& 1.76M$_{\odot}$)
from Type II supernovae (SNe), with a reduced mass for NS resulting from Type Ib SNe (peaking at 1.32M$_{\odot}$) although, as noted by the authors, significant uncertainties in the
input physics remain (e.g. pre SNe mass loss rates).

Unambiguous NS mass determinations in X-ray binaries can only be made for {\em eclipsing} binary systems, where the inclination angle is well constrained. In this paper we aim to derive masses for both
components of the massive eclipsing X-ray binary  OAO~1657--415, rendered possible by a combination of  X-ray timing analysis for the NS and radial velocity (RV)
analysis of the mass donor. The relative scarcity of X-ray pulsars, combined with the low geometric probability that an eclipse of
the binary system will be seen, means that only 11 eclipsing X-ray binary pulsars are currently known. Up to this time 7 of these systems have NS mass determinations (e.g.
\citet{mason10}, \citet{valbaker05}, \citet{quaintrell03}). The NS mass resolution of the remaining systems is thus a priority.

\begin{figure}
\centering
    \includegraphics[scale=0.20]{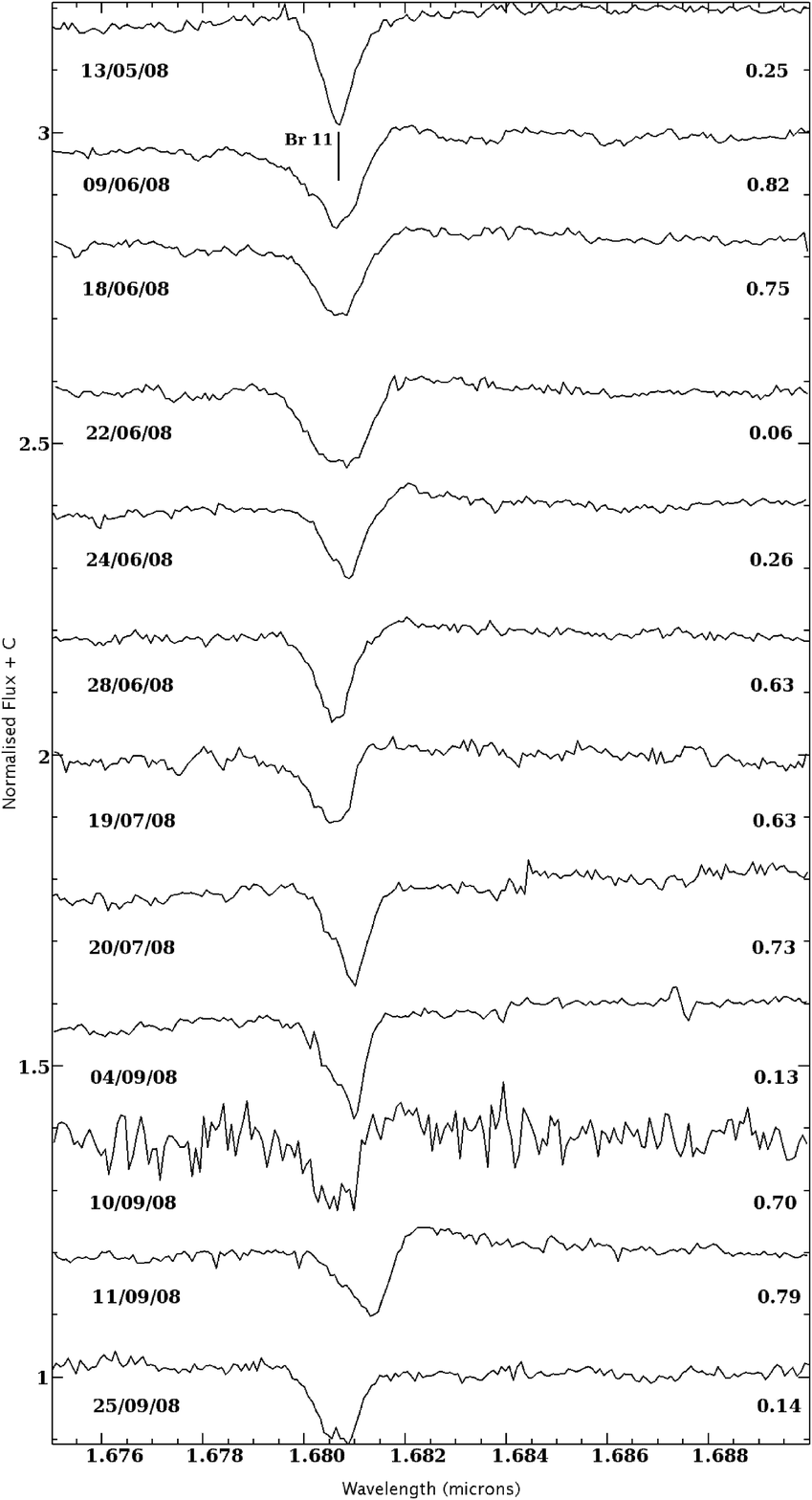}
    \caption{Continuum normalised H band spectra of OAO~1657--415 (highlighting the Brackett 11 line) in date order. The orbital phase corresponding to each spectrum is indicated on the right hand side of the figure.}
\end{figure}

\begin{figure*}
\centering
    \includegraphics[width=14cm]{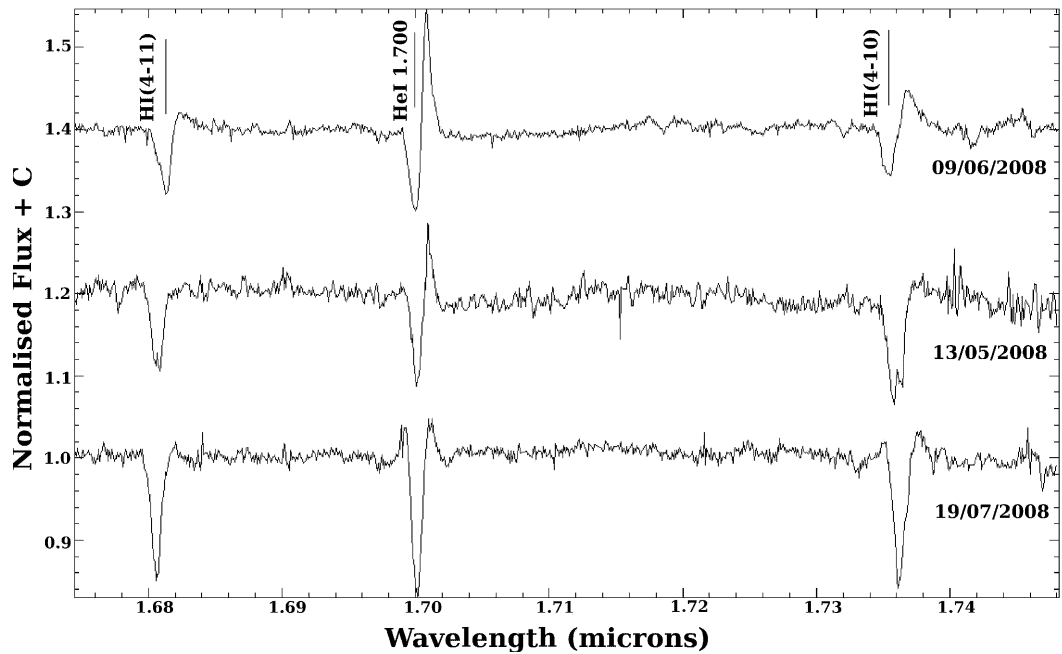}
    \caption{OAO~1657--415 spectral features illustrating the varying wind structure which is particularly noticeable in the P-Cygni line profile of the He {\sc i} 1.700~$\mu$m line.}
\end{figure*}
\subsection{The history of OAO~1657--415}

OAO~1657--415 was first detected over 30 years ago by the {\it Copernicus} X-ray satellite \citep{polidan78}. Follow up observations by {\it HEAO-1} detected pulsations with a period of 38.22~s \citep{white79}, whilst later observations by the {\it Compton Gamma-Ray Observatory} (CGRO) BATSE instrument determined that OAO~1657--415 was an eclipsing accreting pulsar with an orbital period of $\sim 10.44$~days \citep{chak93}.
Along with Vela X-1 \citep{quaintrell03}, OAO~1657--415 is one of only two eclipsing X-ray binary pulsars to exhibit a significant eccentricity, namely $e = 0.107 \pm 0.001$ \citep{jenke11}. It is further differentiated in possessing the longest orbital period (refined by \citet{barnstedt08} as $P = 10.448$~d) of any of these types of eclipsing X-ray binary systems.

From an initial examination of the orbital parameters of the X-ray pulsar it was determined that OAO~1657--415 was a high-mass system, indicating the mass of the donor lay in the range $14-18$~M$_{\odot}$, with a corresponding radius range of $25-32$~R$_{\odot}$. Determination of these stellar parameters led to a spectral classification of B0-6I \citep{chak93}.

The correct identification of the donor in this system required the precise location of OAO~1657--415 to be accurately made. This was achieved by the {\it Chandra X-Ray Observatory} narrowing the X-ray location error radius down to 0.5$^{\prime\prime}$. Optical imaging of this position did not detect any donor candidates down to a magnitude of V$>$23, implying many orders of optical extinction for the presumed massive mass donor. Near infrared imaging was therefore employed, resulting in the identification of a donor located within the {\it Chandra} error radius. A corresponding IR counterpart was located in the {\it 2MASS} catalogue, namely 2MASS J17004888--4139214 with magnitudes J = 14.1, H = 11.7 and K$_{\rm s}$ = 10.4, an extinction of A$_{V} = 20.4 \pm 1.3$, and located at a distance of $6.4 \pm 1.5$~kpc \citep{chak02}.

As a result, near-IR spectroscopy proved mandatory to both classify the mass donor in OAO~1657--415 and also to construct an RV  curve.
Spectroscopy of the donor obtained in 2008 \citep{mason09} led to a re-evaluation of the spectral classification, with the correspondence between published NIR B0-6 spectra \citep{hanson05} and that of OAO~1657--415 proving poor. Close examination revealed that OAO~1657--415 shared a similar spectral morphology with that of Ofpe/WNL stars. These are stars in transition between the OB main sequence and hydrogen depleted Wolf-Rayet stars, whose evolution follows from a wide range of progenitor masses.

\section{Observations and Data Reduction}

As the mass donor in OAO~1657--415 is faint (H $\sim$ 11.7) we employed the NIR spectrometer ISAAC on the VLT to obtain high resolution (R $\sim$ 3000) and high S/N spectra in the H band.
Observations were conducted between 2008 May 13th and 2008 September 25th in the SW MRes  mode with a 0.8$^{\prime\prime}$ slit. Science and RV template exposures centred on 1.7 $\mu$m for a total integration time of 1200s and 128s respectively were obtained.
Spectra were reduced using the ISAAC pipeline and wavelength calibration performed utilising OH skylines.The resulting data has a count rate below 10 000 ADU; therefore no correction for non-linearity was required.

The RV template observed was HD154313, a bright (H $\sim$ 6.8) standard with a known radial velocity of $-29$~km~s$^{-1}$ and close in spectral type (B0Iab) to the previously assumed classification of OAO~1657--415 \citep{chak93}.
Telluric correction was employed to remove atmospheric features from the target and template spectra.

Cross-correlation was performed using the standard IRAF\footnote{IRAF is distributed by the National Optical Astronomy Observatory, which is operated by the Association of Universities for Research in Astronomy, Inc., under cooperative agreement with the National Science Foundation.} routine {\it fxcor}.
Unfortunately due to time constraints and the end of the VLT observing semester we were unable to obtain the full data set of the 20 target spectra we requested. However, 12 high quality spectra that covered a wide range of orbital phase were obtained, sufficient to determine a dynamical mass solution for OAO~1657--415 (Fig. 1).

\section{Data Analysis}

After telluric correction the target spectra can be seen to exhibit three well defined spectral lines: 1.681 $\mu$m Brackett 11 H {\sc i} (4-11), 1.700 $\mu$m He {\sc i} and 1.736 $\mu$m Brackett 10 H {\sc i} (4-10) (Fig. 2). As noted earlier, OAO~1657--415 has been spectrally classified as an Ofpe/WNL star. These transitional stars are characterised by exceptionally intense stellar winds with low terminal velocities and high mass loss rates \citep{martins07}. This is demonstrated in the morphology of the three spectra from the dataset obtained at differing orbital phases (Fig. 2), in which P-Cygni profiles can be seen in each of the spectral lines, the intensity of which clearly varies throughout an orbital cycle.

Stellar wind contamination leading to varying line-profiles make the process of cross-correlation to determine radial velocities problematic. After close examination of each spectrum within the dataset it was decided to cross-correlate around the region surrounding the Brackett 11 1.681~$\mu$m line. We determined that the measured radial velocities of this line would be less affected by the stellar wind, originating closer to the photosphere and exhibiting less P-Cygni profile variation.
Unfortunately for two of the science spectra (\#1 and \#12 in Table 1) RV template observations were not completed. Additionally 2 science spectra (\#3 and \#9 in Table 1) exhibited significantly broadened line-profiles in comparison to other spectra in the dataset. To achieve as accurate an RV measurement as possible it was decided to use the science spectrum \#2 as a template reference spectrum in these four cases. We determined this spectrum possessed the most accurate telluric division and the narrowest line profile.
The resulting radial velocities were then corrected to the solar system barycentre and are shown in Table 1.

\begin{table*}
\caption{The radial velocities of OAO~1657--415 determined from the Brackett 11 absorption line.}
\label{usable_spectra}
\centering
\begin{tabular} {ccccccc}
\hline
Spectrum number & Mid-point of Observations (UT) & MJD & Phase & True anomaly  & Radial velocity / km s$^{-1}$ & RV Std \\ \hline
1 & 2008 May 13.340 & 54 599.41 & 0.249 & 0.283 & --45.77 $\pm$ 10.3  & Reference \\
2 & 2008 Jun 09.022 & 54 626.22 & 0.815 & 0.783 & --99.05 $\pm$ 10.3  & HD154313  \\
3 & 2008 Jun 18.928 & 54 636.00 & 0.751 & 0.718 & --79.26 $\pm$ 10.3 & Reference \\
4 & 2008 Jun 22.238 & 54 639.24 & 0.061 & 0.075 & --39.93 $\pm$ 10.3 & HD154313 \\
5 & 2008 Jun 24.319 & 54 641.32 & 0.260 & 0.293 & --26.02  $\pm$ 10.3 & HD154313 \\
6 & 2008 Jun 28.149 & 54 645.15 & 0.627 & 0.605 & --65.87  $\pm$ 10.3 & HD154313 \\
7 & 2008 Jul 19.115 & 54 666.06 & 0.628 & 0.606 & --66.72  $\pm$ 10.3 & HD154313 \\
8 & 2008 Jul 20.076 & 54 667.08 & 0.725 & 0.694 & --61.13 $\pm$ 10.3 & HD154313  \\
9 & 2008 Sept 04.066 & 54 713.07 & 0.127 & 0.153 & --38.67 $\pm$ 10.3 & Reference  \\
10 & 2008 Sept 10.054 & 54 719.05 & 0.700 & 0.670 & --79.50 $\pm$ 10.3 & HD154313 \\
11 & 2008 Sept 11.013 & 54 720.01 & 0.792 & 0.759 & --81.77 $\pm$ 10.3 & HD154313 \\
12 & 2008 Sept 25.110 & 54 734.11 & 0.141 & 0.169 & --45.52 $\pm$ 10.3 & Reference \\
\end{tabular}
\end{table*}

\section{OAO~1657--415 RV curve fitting}

From X-ray data it has been deduced that OAO~1657--415 has an elliptical orbit with $e = 0.107 \pm 0.001$ \citep{jenke11} and the linear ephemeris of \citet{barnstedt08} gives the epoch at which the mean longitude is $90^\circ$ as
\begin{equation}
T_{90} / {\rm MJD} = 52663.893(10) + 10.44812(13) N
\end{equation}
where the uncertainties refer to the last decimal place quoted and $N$ is the cycle number. The more recent quadratic ephemeris derived by \citet{jenke11} concludes that the orbital period is currently decreasing with $\dot{P}_{\rm orb}/P_{\rm orb} = (-3.34 \pm 0.14) \times 10^{-6}~{\rm yr}^{-1}$, but otherwise is very little different from this. Given the relatively low precision of our radial velocity measurements, the linear ephemeris is perfectly adequate for our purposes. During the period in which our observations were conducted the accumulated uncertainty in phase was calculated as $\sim$ 0.003.

As OAO~1657--415 has an appreciable eccentricity and thus will significantly deviate from sinusoidal motion, we cannot fit the radial velocities of the donor in this system with a sinusoidal radial velocity curve. For an eccentric system we can construct a sinusoidal radial velocity curve by fitting the true anomaly $\nu$ (the angle between the major axis and a line from the star to the focus of the ellipse) to the calculated radial velocities of the donor. In order to do this we must first deduce the mean anomaly $M$ (the time since the last periapsis multiplied by the orbital phase). This in turn is related to the eccentric anomaly $E$ (the angle between the major axis of the ellipse and the line joining the position of the object and the centre of the ellipse) by the equation
\begin{equation}
   E - e \sin E = M = \frac{2\pi}{P} (t-T_0)
\end{equation}
The right hand side of the above equation is the orbital phase, with $T_0$ the time of periastron passage. In an eccentric orbit, $T_0$ and $T_{90}$ are related by
\begin{equation}
    T_0 = T_{90} + \frac{P(\omega - \pi/2)}{2 \pi}
\end{equation}
In the case of OAO~1657--415, the longitude of periastron, $\omega = 91.7^\circ \pm 0.5^\circ$ \citep{jenke11}, so the offset between the two is only about 0.005 in orbital phase. 

This eccentric anomaly equation cannot be solved analytically, but solving it numerically we obtained the eccentric anomaly for each observation. These eccentric anomaly values can then be finally related to the true anomaly by the equation
\begin{equation}
    \tan \left (\frac{\nu}{2}\right) = \left({\frac{1+e}{1-e}}\right)^{\frac{1}{2}} \tan \left(\frac{E}{2}\right)
\end{equation}
The calculated true anomaly values are shown in Table 1 and the measured radial velocities plotted as a function of the calculated true anomaly are shown in Figure 3. Over-plotted on these data are two sinusoidal fits. The first has just two free parameters (RV amplitude and systemic velocity) with a period and zero phase fixed by the ephemeris of Barnstedt et al. (2008). The second fit allows the zero phase as a third free parameter to allow for any possible accumulated uncertainty. However, this is only a marginally better fit and the reduction in chi-squared does not justify the inclusion of an extra free parameter. We therefore use the first fit which yields a radial velocity amplitude of $22.1~\pm~3.5$~km~s$^{-1}$ and a systemic velocity of $57.2~\pm~3.0$~km~s$^{-1}$. Error bars on each radial velocity value shown in Figure 3 are such that the reduced chi-squared of the fit is equal to unity.

\begin{figure}
\centering
    \includegraphics[width=9cm]{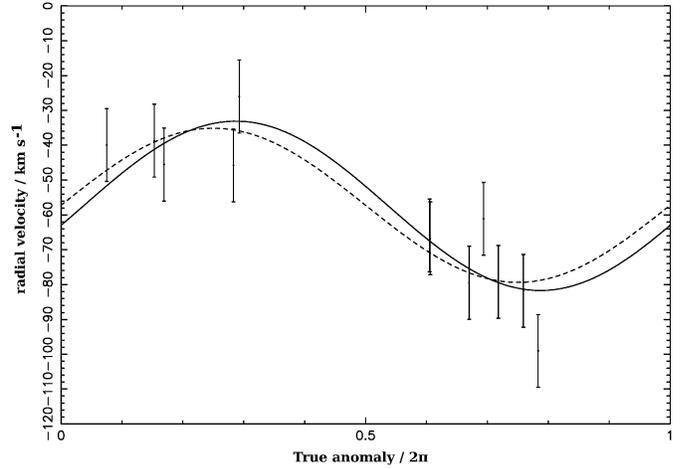}
    \caption{Radial velocity solution for the mass donor in OAO~1657--415. The dashed line represents a fixed zero phase coincident with the published ephemeris (\citet{barnstedt08}). The solid line is the best fitting sinusoid with three free parameters, allowing for a shift in phase.}
\end{figure}

\section{Calculation of system parameters}

Jenke et al (2011) measured the projected semi-major axis of the neutron star's orbit from X-ray pulse timing delays as $a_{\rm x} \sin i = 106.16 \pm 0.08$~light seconds. From this, the semi-amplitude of the neutron star's radial velocity may be calculated using
\begin{equation}
    a_{\rm X} \sin i = \left(\frac{P}{2\pi} \right) K_{\rm X} (1-e^2)^{1/2}
\end{equation}
to give $K_{\rm X} = 222.8 \pm 0.2$~km~s$^{-1}$.

To determine the masses of the system components we must first consider the mass ratio of the system $q$ which is equal to the ratio of the semi-amplitudes of the radial velocities for each star
\begin{equation}
     q = \frac{M_{\rm X}}{M_{\rm O}} = \frac{K_{\rm O}}{K_{\rm X}}
\end{equation}
where $M_{\rm O}$ and $M_{\rm X}$ are the masses of the supergiant star and neutron star respectively, with $K_{\rm O}$ and $K_{\rm X}$ the corresponding semi-amplitudes of their radial velocities. Specifically for elliptical orbits,
\begin{equation}
M_{\rm O} = \frac{{K_{\rm X}}^3 P\left(1-{e}\right)^{\frac{3}{2}}}{2\pi G \sin^3 i}\left(1+q\right)^2
\end{equation}
and similarly
\begin{equation}
M_{\rm X} = \frac{{K_{\rm O}}^3 P\left(1-{e}\right)^{\frac{3}{2}}}{2 \pi G \sin^3 i}\left(1+\frac{1}{q} \right)^2
\end{equation}
where $i$ is the angle between the normal to the plane of the orbit and the line-of-sight and $P$ is the orbital period.

The geometry of the eclipse yields the following relation between the angle of inclination of the system, $i$, the eclipse half angle, $\theta$, and various length scales in the system:
\begin{equation}
     \sin i \approx \frac{\left[1 - \beta^2 \left(\frac{R_L}{r} \right)^2\right]^{1/2}}{\cos~\theta_{\rm e}}
\end{equation}
Here $R_{\rm L}$ is the Roche lobe radius of the supergiant, $r$ is the separation between the centres of the two stars and $\beta$ is the ratio of the supergiant's radius to that of its Roche lobe, referred to as the filling factor. For stars in a circular orbit, $R_{\rm L}$ and $\beta$ are constant, and the separation between the centres of the two stars is just the radius of the orbit. An approximation of the Roche lobe radius is given by \citet{joss84} as
\begin{equation}
   \frac{R_{\rm L}}{r} \approx A + B \log q + C \log^2 q
\end{equation}
where they determine the constants as
\begin{equation}
A \approx 0.398 - 0.026\Omega^2 + 0.004\Omega^3
\end{equation}
\begin{equation}
B \approx - 0.264 + 0.052\Omega^2 - 0.015\Omega^3
\end{equation}
\begin{equation}
C \approx - 0.023 - 0.005\Omega^2
\end{equation}
$\Omega$ is the ratio of the spin period of the supergiant to its orbital period. Since the system has not
circularized, $\Omega$ is unlikely to be unity. If tides are efficient, then we might suppose that the rotation of the donor is synchronized at periastron, where the tides are most effective. In this case, the star's angular velocity will be larger than the orbital velocity at mid-eclipse, and at this point $\Omega \sim 1.2$. To account for this uncertainty in our calculations we therefore adopt $\Omega = 1.2 \pm 0.2$ as a conservative estimate.

For systems with a significantly eccentric orbit such as OAO1657--415, the Roche lobe radius, the separation between the centres of mass of the two stars, and the Roche lobe filling factor will all vary with orbital phase, although $R_{\rm L}$/$r$ can be assumed to be constant. Using the above, we therefore calculate $R_{\rm L}/r = 0.553 \pm 0.025$.

The remaining system parameters were then calculated under two assumptions. If the donor star fills its Roche
lobe at periastron, then the filling factor at mid-eclipse will be  $\beta \sim 0.9$. This allows us to calculate a lower limit to the system inclination, $i$, and upper limits are arrived at for the stellar masses. Conversely, if the system is viewed edge-on ($i=90^\circ$), we calculate a lower limit for the Roche lobe filling factor, $\beta$ at mid-eclipse, and lower limits are derived for the stellar masses. The limits on the semi-major axis of the orbit, $a$, are simply determined from Kepler's third law using the calculated ranges for the masses of the two stars. In practice, it turns out that the uncertainties on the system parameters overlap significantly for these two limits, so in Table 2 we list a single set of values.

Now, the separation between the centres of the two stars at mid-eclipse is given by
\begin{equation}
    r = \frac{a(1-e^2)}{1+e \cos \omega}
\end{equation}
where $\omega$ is the argument of periastron, measured by Jenke et al (2011) to be $\omega = 91.7^{\circ} \pm 0.5^{\circ}$ in this case. This yields a value of $r = 49.9 \pm 1.0$~R$_{\rm \odot}$. Then the Roche lobe radius, $R_{\rm L}$, is this separation multiplied by the factor $0.553 \pm 0.025$ calculated earlier, which gives the value listed in Table 2. Finally, the radius of the donor star, $R_{\rm O}$ is just $\beta R_{\rm L}$, and this value is also shown in Table 2.

Uncertainties on each of the inferred system parameters shown in Table 2 were determined by a Monte-Carlo calculation which propagated the uncertainties in each observed parameter through one million trials, assuming a Gaussian distribution in each individual uncertainty.

\begin{table}
\caption{System parameters for OAO~1657--415.}
 \label{results}
 \begin{tabular}{lll} \hline
Parameter                    & Value & Ref. \\ \hline
{\it Observed}              &   &    \\
$a_{\rm X} \sin i$ / lt sec & $106.16 \pm 0.08$          & [1]\\
$P$ / d                     & $10.44812 \pm 0.00013$   & [2]\\
$T_{90}$ / MJD              & $52663.893 \pm 0.001$    & [2]\\
$\omega$                    & $91.7 \pm 0.5$               & [1] \\
$e$                         & $0.107 \pm 0.001$        & [1]\\
$\theta_{\rm e}$ / deg      & $29.7 \pm 1.3$           & [3]\\
$K_{\rm O}$ / km s$^{-1}$   & $22.1 \pm 3.5$           & [4]\\
{\it Assumed}               &   &   \\
$\Omega$                    & $1.2 \pm 0.2$             &  \\
{\it Inferred}              &   &   \\
$K_{\rm X}$ / km s$^{-1}$   & $222.8 \pm 0.2$         & \\
$q$                         & $0.099 \pm 0.016$        & \\
$\beta$                     & $0.896 \pm 0.051$                & \\
$i$ / deg                   & $87.0 \pm 7.7$              & \\
$M_{\rm X}$ / M$_{\odot}$   & $1.42 \pm 0.26$      & \\
$M_{\rm O}$ / M$_{\odot}$   & $14.3 \pm 0.8$             & \\
$a$ / R$_{\odot}$           & $50.3 \pm 1.0$               & \\
$R_{\rm L}$ / R$_{\odot}$   & $27.6 \pm 1.6$             & \\
$R_{\rm O}$ / R$_{\odot}$   & $24.8 \pm 1.5$            & \\ \hline
\end{tabular}\\
$[1]$ Jenke et al. 2011; $[2]$ Barnstedt et al. 2008;\\
$[3]$ Chakrabarty et al. 1993; $[4]$ this paper
\end{table}

\section{Discussion}

\subsection{System properties}

 The H band spectra for the period May - July, 2008 (Fig. 2) demonstrate a high degree of line profile
variability (LPV) in the H\,{\sc i} transitions,  with the H\,{\sc i} (4-10) line observed to transition from absorption to  a P Cygni profile. He {\sc i} 1.70 $\mu$m is also present and displays a variable P Cygni profile, although He {\sc ii} 1.69$\mu$m  is absent. Conversely, the K band spectrum of this object previously  published  (Fig 2. \citet{mason09}), presents the He\,{\sc ii} 2.189 $\mu$m line in emission. The H band  spectrum is  more reminiscent of P Cygni early B  hypergiants than the Ofpe/WNL classification that had been inferred from  the K band data alone. Indeed, the H and K spectra can only be reconciled if it is assumed that the He\,{\sc ii} 2.189  $\mu$m transition is driven into  emission by the presence of the neutron star.

Precedents for excess  He\,{\sc ii} emission in HMXBs in both the optical and near-IR are provided by LMC
X-3 \& X-4 \citep{nc} and  CI Cam \citep{clark99} respectively. Radial velocity shifts in this line in
LMC X-4  confirm its association  with the compact accretor, with the  observed line strength variability
possibly resulting from changes in  accretion rate; similar observations for OAO~1657--415 would test  the hypothesis that the line is likewise associated with the accretor rather than the primary. The significant rapid, non-secular LPV in the remaining H\,{\sc i} and He\,{\sc i} lines is most easily understood by a departure from spherical symmetry by the presence of the neutron star. The  WNLh star Wd1-44 provides a precedent for this conclusion, also demonstrating dramatic LPV in wind dominated  H\,{\sc i} and He\,{\sc i} P Cygni profiles, although in this case it is expected that the companion is a normal H/He-burning object \citep{clark10}.

Mindful of the above, we attempted to derive the physical and wind properties of the donor star in OAO 1657--415 using the stellar  atmosphere modelling code CMFGEN \citep{hillier98} and the method as described in Section 4.1 of \citet{crowther09},  utilising the H and K band spectra obtained between May - September 2008. Unfortunately, we were unable to obtain a satisfactory formal fit to these
data, which we attribute to the reasons above, namely a lack of contemporaneous data for  a star showing LPV -  possibly as  a consequence of an aspherical wind - and the presence of an additional source of ionising radiation within the system yielding discrepant   He {\sc ii} 2.189 $\mu$m emission.

While we were unable to obtain a formal fit to the model, we were able to determine
 order of magnitude estimates for the bulk properties of the star that we regard as
indicative of its true properties: $T~\sim 20$~kK,
$L \sim 10^{5.1}$L$_{\odot}$, $R \sim 30$~R$_{\odot}$, $\dot{M} = 2 \times 10^{-6}$~M$_{\odot}$~yr$^{-1}$,
$v_{\infty} \sim 250$~km~s$^{-1}$. Using our spectra from the H and K bands the model also provided us with estimates of the mass fraction of hydrogen {\it X}$_{H}$ = 5\% (H/He = 0.21 by number) in which our primary diagnostics were the Brackett series lines (1.680, 1.736 and 2.166 $\mu$m), helium {\it X}$_{He}$ = 94\% by mass from He {\sc i} lines (1.700, 2.058, 2.112, 2.161 $\mu$m) and other metals {\it X}$_{Z}$ = 1\%.

Comparing the stellar parameters of OAO~1657--415 with the Ofpe/WNL star AF \citep{martins07}, we find a correspondence between temperature and radius, whilst the luminosity of OAO~1657--415 ($L \sim 10^{5.1}$L$_{\odot}$) is consistent within the uncertainties to that of AF.     
However, the lower than expected luminosity of OAO~1657-415 implies that it could be a B supergiant rather than a Ofpe/WNL star. Noting that there is a large degree of overlap between model derived luminosities of Ofpe/WNL and B supergiants, we turn to other stellar properties to address this possibility. We find the low terminal velocity of OAO~1657--415 combined with its mass loss rate and temperature are incompatible with those found for field Blue Supergiants \citep{bsg}. B supergiants with a luminosity of $L \sim 10^{5.1}$L$_{\odot}$ would be expected to have a radius greatly in excess of that found from both modelling and dynamical calculations of $\sim$ 30 M$_\odot$ \citep{searle08}. B supergiants also have a significantly greater hydrogen abundance of H/He = 5 by number \citep{bsg} than that found during our modelling of OAO~1657--415. With the degree to which hydrogen is depleted within OAO 1657-415 its radius may seem excessively large. However, several examples of WNL stars with atmospheres with a similar or greater degree of hydrogen depletion and comparable radii to that of OAO 1657-415 have been detailed in the literature. In particular, the emission line star HD 326823 \citep{marcolino07} is reported as entering the Wolf-Rayet phase as a WN8 star with a radius R = 30 R$_{\odot}$ and a hydrogen mass fraction of {\it X}$_{H}$ $\sim$ 3\%. Similarly the WN8 star WR123 \citep{crowther95} may have a hydrogen mass faction as low as 0.5\% yet is reported as having a radius R $\sim$ 15 R$_{\odot}$. 


\subsection{OAO~1657--415 in an evolutionary context}

When these properties are combined with the finding that OAO~1657--415
possesses a severely H-depleted atmosphere, it is apparent that it is a rather evolved object and is clearly more advanced  than the  typical OB SG mass donors in HMXBs.
A comparable mass donor in this regard is Wray 977, the blue hypergiant companion within GX301-2. Prior to OAO 1657--415, Wray 977 was the most evolved non-WR donor known within a HMXB system. Wray 977 has a significantly higher hydrogen mass fraction of 45\%, (Table 8 in \citet{kaper06}) than that of OAO1657--415.

Thus, while a definitive spectral classification for the mass donor in OAO~1657--415 is difficult,
even if it has yet to formally become a WR star, it is apparent that it will do so shortly. As such
it will become only the fourth known HMXB with such an evolved mass donor. The only other example in our Galaxy is Cyg X-3, which has a WNE mass donor orbiting a NS or BH companion in a 4.8~hr orbit \citep{vk}. \citet{vdh} describe an evolutionary scenario for this system, whereby the mass donor in the OB SG+accretor progenitor system expands towards a RSG phase, triggering  Common Envelope evolution which leads to the formation of the WNE star and a dramatic contraction in orbital separation.

In contrast, while  the two known extragalactic examples - IC10 X-1 \citep{clark04} and NGC300 X-1
\citep{crowther07} - also host WN stars, they both have significantly longer orbital periods ($\sim 30$~hrs), and appear to host more luminous mass donors and more massive accretors\footnote{$>15$M$_{\odot}$ for NGC300 X-1 (\citep{crowther10}) and $>23$M$_{\odot}$ for IC10 X-1 (\citep{silverman}) versus $<10$M$_{\odot}$ for Cyg X-3 (\citep{hanson}).}. These properties preclude these binaries from evolving via the scheme inferred for Cyg X-3. Instead, it is thought that they must have evolved from very compact initial O+O star binaries  via  the case M pathway described by \citet{demink09}. In this, the rapid rotation of both (tidally locked) components leads to efficient mixing and hence homogeneous chemical evolution, which results in the stars remaining  within their Roche lobes while evolving to WR stars. This evolution prevents binary driven mass loss, resulting in a longer final orbital period and crucially a heavier relativistic companion.

However, the properties of OAO~1657-415 do not fit either of these schemes. The non-zero
eccentricity appears to preclude post-SN common envelope evolution (as a result of
circularization processes during spiral-in) and hence OAO~1657-415
being the endpoint of the van~den~Heuvel \& De~Loore~(1973) pathway. To address this in a more
quantitative manner we adopt the formalism of \citet{dewi00}, whereby the
evolution of a putative RSG+NS system through a common envelope (CE) phase may be followed by
the paramerisation of the binding energy of the envelope of the donor star ($\lambda$) and the
efficiency parameter $\eta$ for the removal of the envelope during spiral-in. In the following
discussion we assumed an initial donor mass of $\sim\!40\,M_{\odot}$ resulting in a
$\sim\!24.5\,M_{\odot}$ envelope and $\sim\!15.5\,M_{\odot}$ core, representative of the
current mass of the Ofpe/WNL star. The binding energy of the envelope of the mass donor
is given by
\begin{equation}
\label{Ebind}
   E_{\rm bind} = \frac{-GM_{\rm donor}M_{\rm env}}{\lambda R_{\rm donor}}
\end{equation}
where $M_{\rm donor}$, $M_{\rm env}$ and $R_{\rm donor}$ are the masses of the donor
and envelope of the donor and the donor radius, respectively. In order to facilitate
a successful ejection of the donor star envelope we consider a wide orbit prior to the
CE for two reasons: A wide orbit requires an evolved donor star to fill its Roche-lobe
and giant stars have relatively small binding energies of their envelope. Furthermore,
a wide pre-CE orbit has the potential to release more orbital energy during spiral-in.
Even for a rather evolved donor ($R_{\rm donor} \simeq 1500\,R_{\odot}$) we find
$|E_{\rm bind}| \sim 10^{50}$~erg given that for such a star $\lambda \le 0.02$
(c.f. \citet{dewi01}). However, simply determining the available orbital energy:
\begin{equation}
\label{Ebind2}
   \Delta E_{\rm orb} = \frac{GM_{\rm core}M_{\rm NS}}{2a} - \frac{GM_{\rm donor}M_{\rm NS}}{2a_0}
                      = 8.2\times10^{47} \rm{erg}
\end{equation}
where the mass of the neutron star, $M_{\rm NS}=1.5\,M_{\odot}$ and $a_0$ and $a$ are the
initial and final orbital separations.
Thus even assuming the smallest possible envelope binding energy and a 100\% ejection
efficiency ($\eta =1.0$) for conversion of orbital energy we find that there is
insufficient energy available for ejection of the donor envelope (by 2 orders of
magnitude). This is indicated graphically in Fig.~4 which demonstrates that for
a successful ejection of the envelope yielding the current orbital parameters
one would require $\lambda > 1$ (solutions are only possible above the dashed line),
while stellar structure calculations for a
$40\,M_{\odot}$ star yield $\lambda = 0.006 - 0.02$ (Dewi \& Tauris~2001).

If  the common envelope evolutionary pathway of \citet{vdh} cannot produce OAO1657--415 what might?
The large  degree of H-depletion in the mass donor  implies a
post-rather than pre-RSG state; hence  it is difficult to interpret  the system as  evolving {\em towards} such a common envelope phase.  Moreover, case M chemically homogeneous evolution (e.g. \cite{demink09}) also seems inappropriate since it would result in a much more massive relativistic companion  than we find.

In the absence of any tailored numerical model for
 the evolution of OAO~1657--415 (c.f. \cite{well} for GX301-2) any scenario is inevitably
somewhat speculative.  It might be supposed that  the progenitor system consisted of two O stars
in a comparatively short orbit. Mass is transferred from the primary to the secondary, which after
 rapidly gaining mass   evolves more quickly than the donor and hence undergoes
SNe first, leaving the primary  in a H depleted phase. Despite such mass transfer leading
 to a rather wide pre-SN binary configuration, a fortuitous SN kick may reduce the separation, with subsequent interaction acting to circularise the orbit. While this would leave the primary in the requisite H-depleted state, simulations of  conservative mass transfer (e.g. \cite{wlb})  leading to  reverse SNe yield very low masses for the mass donor in comparison to our dynamical mass estimate.
Alternatively, from our earlier paper on OAO1657--415, \citep{mason09} luminosity calculations and comparison to evolutionary tracks for massive stars from \citet{meynet00} indicate a progenitor mass for the OAO1657--415 donor of $\sim$ $40M_{\odot}$. This implies OAO1657--415 is massive enough to have undergone an LBV phase \citep{langer94}. If the evolution of the progenitor system had proceeded such that a LBV+NS was produced with an orbital period in which both components would interact during the LBV phase, in this case the binding energy of the donors envelope would be reduced whilst the corresponding efficiency parameter $\eta$ for the removal of the envelope would increase. The changes to these parameters might be sufficient to enable the removal of the donors envelope by common envelope evolution. This scenario has been suggested qualitatively by \citet{dedonder97}. The hydrogen depleted HMXB system GX301-2 which hosts a blue hypergiant donor of $43M_{\odot}$ $\pm$ 10$M_{\odot}$  with a 1.85 $\pm$ $08M_{\odot}$ NS companion in an eccentric 41d orbit may provide an analogue to the progenitor system to OAO1657--415. Modelling of this system by \citet{wellstein99} as a $25+26M_{\odot}$ binary which undergoes conservative case A mass transfer with an initial 3.5 day period was able to reproduce the current observed mass of the donor successfully, demonstrating that it is possible for a massive hypergiant to have a NS companion.

We are thus currently unable to determine a definitive qualitative evolutionary pathway for OAO~1657--415, but conclude that it cannot have formed via either of the scenarios proposed for the other three known WR X-ray binaries -   consequently this class of object appear to be rather heterogeneous in nature
in comparison to other subsets of HMXB such as the OB SG and Be star X-ray binaries.

However we do note that the future evolution of the mass donor in  OAO~1657--415 will result in an increase in wind velocity, while wind driven mass loss will cause the separation to increase. The combination of both effects will result in a reduction in accretion efficiency and hence X-ray luminosity. Hence one might suppose that even if the evolutionary pathway that formed OAO1657--415 is traversed by large numbers of stars, comparatively few will be detectable as X-ray sources due to the likely rapidity of passage through the Ofpe/WNL phase.

\subsection{Future evolution of OAO~1657-415}
The final mass of the Wolf-Rayet star prior to its core-collapse is probably somewhere
between $5-10\,M_{\odot}$ \citep{Meynet05}. If the star loses its mass in the form of a direct fast wind,
(assuming the vast majority of the mass is lost from the binary) the system will
have an orbital period of more than 65~days when the companion is reduced to $5\,M_{\odot}$ prior to
the collapse in less than 1~Myr. If the outcome of the core collapse is a neutron star then the ejection of the envelope
surrounding the iron core is likely to disrupt the binary; a binary
will be disrupted if it loses more than half of its total mass in an instantaneous event like a supernova \citep{colgate70}.
On the other hand, the momentum kick expected to be imparted to the newborn neutron star \citep{lyne94}
can also, depending on its direction, have the effect of tightening the binary so it survives and remains bound
despite losing more than half of its total mass \citep{hills83}. In this case a double neutron star system is formed and would most likely be wider than the widest of the currently 9 double neutron star systems known in our Galaxy,
PSR~J1811-1736, which has an orbital period of 18.8~days.\\


\begin{figure}
\centering
    \includegraphics[angle=270,width=9cm]{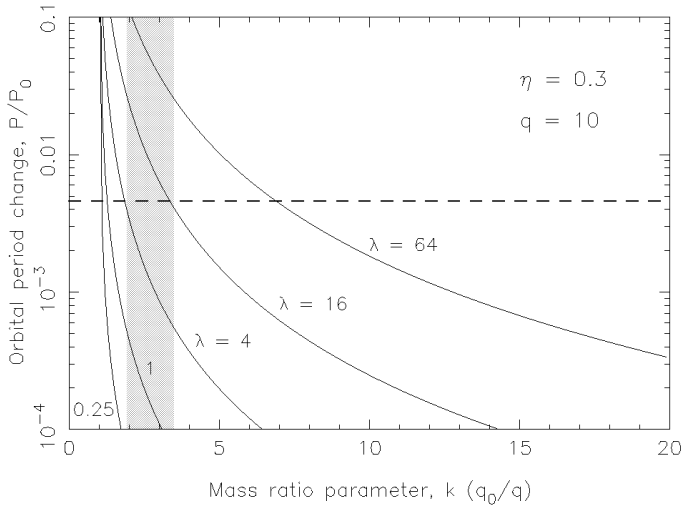}
    \caption{Plot of the initial/final mass ratio (in this case $q$ is defined as $M_{\rm core}$/$M_{\rm NS}$), $k (=q_0/q)$, versus the change in the orbital period - $P/P_0$. The dashed line represents the value of the latter assuming a maximum orbital separation of $\sim 2500$~R$_{\odot}$ ($P_0=2250$~days), while the shaded region corresponds to the range valid for $k$. The curved lines
represent different binding energies of the envelope of the donor star ($\lambda$), assuming $q=10$ and an ejection efficiency $\eta=0.3$. See Sect 6.2}
\end{figure}

\section{Conclusions}

Following on from the recent spectral classification of the High Mass eclipsing X-ray binary pulsar OAO~1657--415 \citep{mason09}, we have reported the analysis and results from multi-epoch observations of the object using the ESO/VLT NIR spectrograph ISAAC. We have constructed an orbital solution from the measured radial velocities which, combined with the orbital parameters of this eccentric system, provides a dynamically determined neutron star mass of $1.42 \pm 0.26$~M$_{\odot}$. The donor star in this system was previously determined to be of type Ofpe/WN9, that has experienced a significant degree of mass loss. The radial velocity curve constructed showed that the donor mass lies within the range $14.3 \pm 0.8$~M$_{\odot}$.

An attempt was made to model the stellar atmosphere of the donor; this was found to be problematic due to the presence of He {\sc ii} emission that made a formal fit to the spectrum difficult. Modelling of the stellar atmosphere did however provide order of magnitude estimates of stellar parameters, indicating that the Ofpe/WN9 star in this system, although significantly hydrogen depleted, has retained a small fraction of its envelope. This combined with the non-zero eccentricity and $\sim 10$~day orbital period of the system is indicative of a different evolutionary pathway than in the three other WR-CC systems known. Calculating the binding energy of the envelope we found there is insufficient energy to eject the envelope. This combined with the orbital period of the system effectively rules out Common Envelope evolution as a viable evolutionary scenario. We have postulated other evolutionary pathways for OAO~1657--415 in which mass transfer occurs in a short binary orbit in which the secondary is the first to experience a SNe, or the system experiences a 'mild' form of CE evolution during an LBV phase. Whilst we are unsure of the exact evolutionary process responsible for the current configuration of OAO~1657--415, we are convinced that the evolution of this system forms an entirely new evolutionary scenario than the two proposed to explain the development of the other three known Wolf-Rayet X-binary systems.

\section*{Acknowledgements}
ABM acknowledges support from an STFC studentship. JSC acknowledges support from an RCUK fellowship.
This research is partially supported by grants AYA2008-06166-C03-03 and
Consolider-GTC CSD-2006-00070 from the Spanish Ministerio de Ciencia e
Innovaci\'on (MICINN). Based on observations carried out at the European Southern Observatory, Chile through programme ID 081.D-0073(B). We thank the anonymous referee for their constructive comments which have helped to improve this paper.

\bibliographystyle{mn2e}
\bibliography{fourth_paper}

\end{document}